\documentclass[conference]{IEEEtran}
\IEEEoverridecommandlockouts

\usepackage{cite}
\usepackage{amsmath,amssymb,amsfonts}
\usepackage{algorithmic}
\usepackage{graphicx}
\usepackage{textcomp}
\usepackage{xcolor}

\def\BibTeX{{\rm B\kern-.05em{\sc i\kern-.025em b}\kern-.08em
    T\kern-.1667em\lower.7ex\hbox{E}\kern-.125emX}}
    
\begin{document}

\title{ARIES: A Scalable Multi-Agent Orchestration Framework for Real-Time Epidemiological Surveillance and Outbreak Monitoring\\
}

\author{\IEEEauthorblockN{Aniket Wattamwar}
\IEEEauthorblockA{\textit{Department of Computer Science} \\
\textit{California State University,}\\
Fullerton, CA \\
orcid.org/0009-0001-1454-838X}
\and
\IEEEauthorblockN{Sampson Akwafuo}
\IEEEauthorblockA{\textit{Department of Computer Science} \\
\textit{California State University,}\\
Fullerton, CA \\
orcid.org/0000-0001-8255-4127}
 }

\maketitle

\begin{abstract}
Global health surveillance is currently facing a challenge of Knowledge Gaps. While general-purpose AI has proliferated, it remains fundamentally unsuited for the high-stakes epidemiological domain due to chronic hallucinations and an inability to navigate specialized data silos. This paper introduces ARIES (Agentic Retrieval Intelligence for Epidemiological Surveillance), a specialized, autonomous multi-agent framework designed to move beyond static, disease-specific dashboards toward a dynamic intelligence ecosystem. Built on a hierarchical command structure, ARIES utilizes GPTs to orchestrate a scalable swarm of sub-agents capable of autonomously querying World Health Organization (WHO), Center for Disease Control and Prevention (CDC), and peer-reviewed research papers. By automating the extraction and logical synthesis of surveillance data, ARIES provides a specialized reasoning that identifies emergent threats and signal divergence in near real-time. This modular architecture proves that a task-specific agentic swarm can outperform generic models, offering a robust, extensible for next-generation outbreak response and global health intelligence.
\end{abstract}


\section{Introduction}
Over the years, lots of data has been published by the World Health Organization (WHO), Centre for Disease Control and Prevention (CDC) \cite{b26}, and National Library of Medicine (NCBI). This data is available in all types of format like XML, JSON, text and tables. Based on this data, there are numerous dashboards for outbreaks, analysis, and trends. Algorithms like time series forecasting, machine learning algorithms are implemented to predict the nature of the disease spread. WHO published Disease Outbreak News (DONs) \cite{b27}, \cite{b28} frequently with information about the assessment, and advice. These dataset are huge in numbers and it is quite an engineering challenge to curate it for a specific task. A significant effort is required to collect, query, analyse and show the data in the form to make decisions and analyse trends. In real time, it is extremely challenging to provide insightful information by querying multiple sources and gathering text in various formats causing high latency and delaying decision and actions.

In recent years, with the development of Large Language Models (LLMs) like GPT by OpenAI \cite{b19}, \cite{b20} it has improved our way of extracting information and data tremendously. LLMs have also been trained on multiple research and literature texts spanning from the most mundane information to all the information available on the internet. LLMs are developed in phase and various LLMs are trained on multiple information sources. The use of LLMs in computational epidemiology and disease outbreak has also increased. Information Extraction of various formats and analyzing it is also another significant engineering challenge \cite{b14}, \cite{b16}.
These LLMs are generalized models that have the ability to answer questions based on the user query. They are also prone to hallucinations and wrong factual information or extracting outdated information. Knowledge Gap exists between the LLM model trained and the latest advancement in a particular field. LLMs work better in one specific domain or area. However, these LLMs have an overhead cost associated with it. Regarding its usage in surveillance and outbreak there is a need to have a framework that would allow users to have conversations \cite{b15} in a way that reduces the risk of hallucination, is specified for that task only and retrieves relevant and updated information only. 

To tackle this challenge, this paper introduced ARIES, which is a multi agent framework for automated disease surveillance. It acts as a Decision Intelligence System with multiple agents performing respective and assigned tasks. The approach is scalable with the current architecture where new data sources emerge then spawning new agents and tools can be incorporated without any redesign.



\section{Related Work}
Samaei et al. \cite{b1}  introduced EpidemIQ which is a multi-agent framework that takes input from users and specific agents to perform tasks like literature review, data visualization, network modeling and expert tasks to create a full manuscript in scientific format. The tool being very powerful is not developed to take into account the recent outbreak news, trends of disease spread. But, this paper introduces that factor of real time outbreak analysis and reasoning using multi agent architecture with state-of-the-art technologies.

Large language models are constantly used for multiple purposes from summarization, extraction, generation, and analysis. Kwok et al. \cite{b2} and Patlolla, Padmavath et al. \cite{b9} and others used similar methods to chat with the LLM like ChatGPT to estimate two important epidemiological parameters i.e reproductive number and epidemic size. With natural language generation and clarifying the objectives and refining the responses using the susceptible-exposed-infected-recovered (SEIR) framework a classic disease transmission model can be utilized. This approach although effective limits us to using tools like ChatGPT which still use generalized Language Models not having entire access to the previous data. For complex transmission models, the natural language generation uses the entire context window. The free version limits users to converse for a limited time in the same context. Changing the context, the user loses all the progress done so far. This prevents generating the models or generating inaccurate models. Adopting ARIES makes sure that data is recent since the agents are using tools from the official APIs or data sources.

R. Allard et al. \cite{b3} showcases time series on disease surveillance with ARIMA (autoregressive, integrated, moving average) algorithms. The paper concluded with the use of forecasting allowed decision makers to consider variability if not detect outbreaks. However, it does not include intelligence in terms of Generative AI. It focuses more on traditional Machine Learning(ML) algorithms. This is similar to the work by Wattamwar et al\cite{b7} and \cite{b8}, where global outbreaks of viral disease are monitored using a dedicated approach. But with ARIES when a user asks a query it can make predictions similar to forecasting based on the data received from the tools associated with the agents.

Before Large Language Models became predominant in most of the domains, machine learning and deep learning played a critical role in finding and detecting patterns. It also meant spending a lot of time in data collection, data cleaning and data manipulation. Syed Ziaur Rahman et al. \cite{b4} and others analyzed multiple algorithms like temporal and spatial prediction models and further with risk prediction models to find the correlation between the disease episodes and the characteristics of it. Deep Neural Networks (DNNs) and long-short term memory (LSTM) models along with Semi-Supervised Learning (SSL) algorithms performed by S. Chae \cite{b10} , Kim, J. \cite{b11}, Abougarair, A. \cite{b12} et al. applied across a wide range of text like social media data, and web articles governed by official bodies show promising results.

Consoli, Sergio et al \cite{b5} and others have performed information extraction (IE) on datasets provided by WHO and Disease Outbreak News. The extraction is carried across multiple LLMs like Llama-2-70b-chat, Mistral-7b-openorca, Mpt-30b-chat, Pythia-12b,  Gpt-4-32k and more. The paper shows results of the accuracy of these models tested across the ProMed and the WHO DONs datasets. This shows the ability of these models to provide accurate results when models are used in ensemble. Sara De Luca \cite{b13} performs zero shot classifications using LLMs with 90.2\% precision on two case studies of Covid19 from 2020-2022 and Mpox Epidemic in Europe. The purpose of the Outbreak monitoring approach introduced \cite{b13} was to detect anomalies in News outbreaks and classify ICD-9-CM disease codes.

Inspired by this, the approach is utilized in the ARIES architecture where each agent LLM can be updated to serve the purpose, thus reducing cost and latency.

\section{Methodology}

ARIES is a hierarchical multi-agent architecture for retrieval of data with multiple agents. The system is designed to navigate through the complexities of multiple global data sources that mimics the operational hierarchy of a public health Emergency Operations Center (EOC).

Unlike sequential processes that risk cumulative error propagation, ARIES utilizes a command-and-control structure where a central "Manager" agent orchestrates a swarm of autonomous sub-agents. In this paper below agents are configured and tested across a set of questions.

\subsection{Manager Agent}\label{AA}
Acts as a Chief Medical Officer. It is responsible for query decomposition, task delegation, and the final synthesis of the response. The Manager evaluates the outputs of the sub-agents for logical consistency and identifies context across sources.

\subsection{Specialized Agents}
A layer of sub-agents, each assigned a domain-specific persona and toolset
\begin{itemize}
\item \textbf{Senior Medical Scientist:} Interfaces with the NCBI PubMed database using the BioC-JSON interoperability format to extract latest literature and research text based on the query.
\item \textbf{CDC Data Analyst:} Automates complex XML-POST requests to the CDC WONDER servers to retrieve data.
\item \textbf{WHO Intelligence Officer:} Monitors the WHO Disease Outbreak News (DONs) via OData-filtered API calls to retrieve real-time risk assessments and international health advisories.
\end{itemize}

\begin{figure}[htbp]
\centerline{\includegraphics[width=\linewidth]{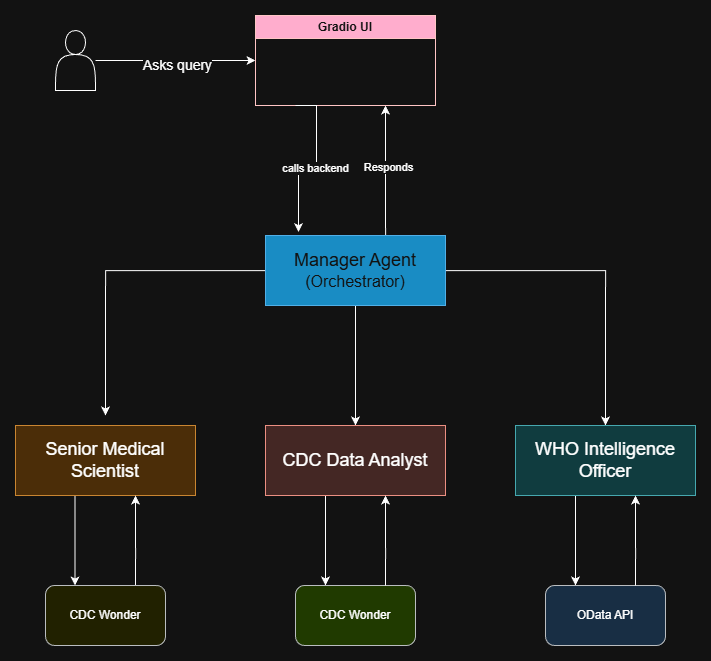}}
\caption{Architecture Diagram}
\label{archifig}
\end{figure}

Figure 1 shows the simple working architecture and responds back with the answer of the user's query. All the agents are divided per task and new agents can be added to it without redesigning the entire system.

\subsection{Operational Execution}
The operational execution of ARIES follows a four-stage "Investigative Loop".

\begin{itemize}
\item \textbf{Ingestion:} The user provides a natural language query (e.g., "Assess the risk of the current avian influenza surge in North American dairy workers").
\item \textbf{Decomposition \& Delegation:} The Manager breaks the query into clinical (PubMed), statistical (CDC), and regulatory (WHO) tasks, dispatching them to the sub-agents in parallel.
\item \textbf{Reasoning:} Sub-agents return their findings to the Manager. The Manager performs Logic Verification, checking for contradictions such as a WHO "Low Risk" alert coinciding with a CDC "Mortality Spike."
\item \textbf{Final Briefing:} The LLM \cite{b17} \cite{b18} used displays the underlying agent logic to the user for full transparency.
\end{itemize}

The architecture for the paper is built using Python and CrewAI. CrewAI framework has the capability to assign a role, goal, and backstory to each agent developed. The agents can also be configured to use any LLM models. If the task is simple a cheaper smaller LLM can be configured to save inference costs and reduce latency. The approach is tested with gpt-4o for all the agents.
The task is also defined with a description to triangulate the three sources once data is fetched and sent to the manager agent.

\subsection{Workflow}

\begin{figure}[htbp]
\centerline{\includegraphics[width=\linewidth]{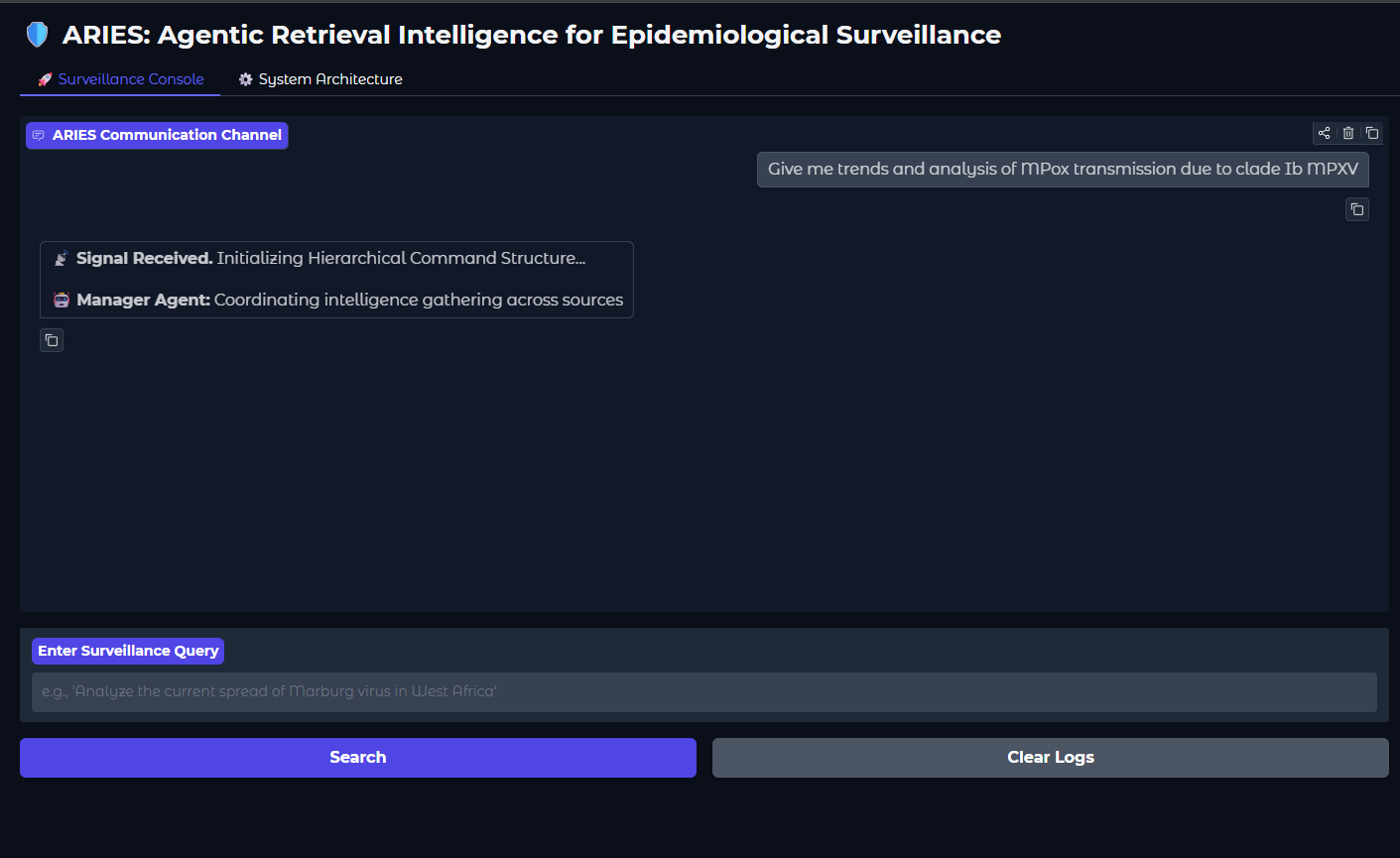}}
\caption{UI of ARIES}
\label{fig2ui}
\end{figure}

Figure 2 shows the UI and how the user interacts with the system. The system shows server side events where the user can see how the model is thinking and what type of flow its following. The query is sent to the backend and the Manager or the Orchestrator agent is triggered who will analyse the query and the intent.

\begin{figure}[htbp]
\centerline{\includegraphics[width=\linewidth]{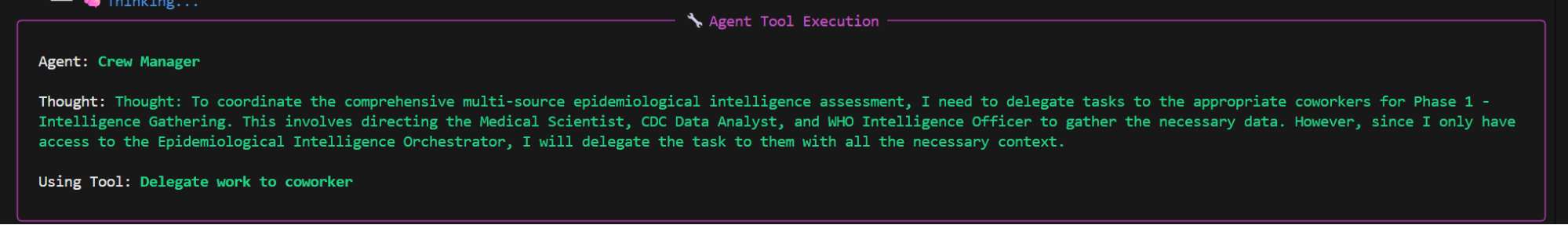}}
\caption{Thought and Intent Identified by Manager}
\label{fig3}
\end{figure}

Figure 3 shows the thought of the agent and understands that it should perform information gathering from the developed agents.

\begin{figure}[htbp]
\centerline{\includegraphics[width=\linewidth]{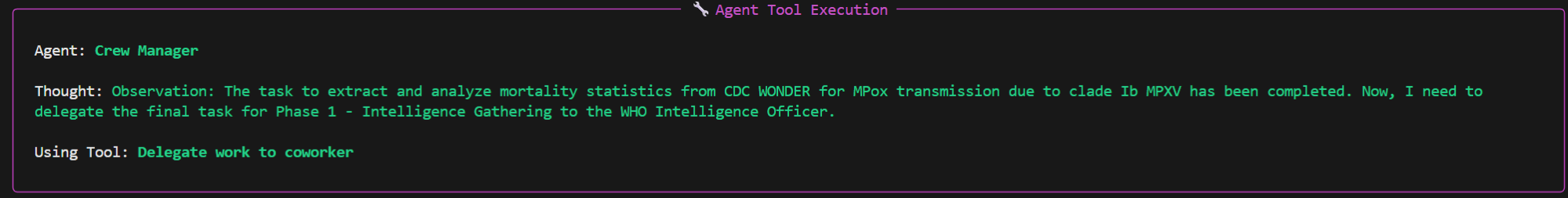}}
\caption{Identifying Agents}
\label{fig4}
\end{figure}

\begin{figure}[htbp]
\centerline{\includegraphics[width=\linewidth]{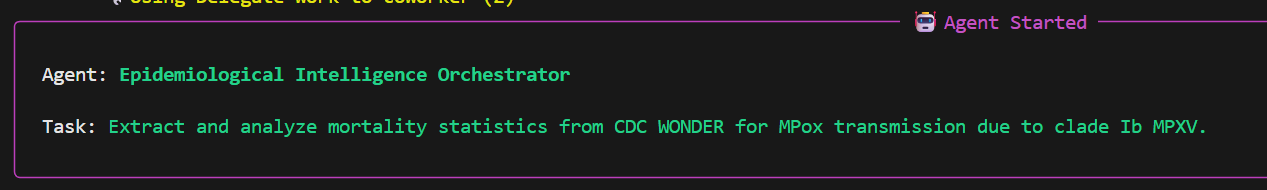}}
\caption{Delegating Task to Sub Agent}
\label{fig5}
\end{figure}

Figure 4 and figure 5 show that task from the orchestrator where it will delegate the work to the CDC Data Analyst and the WHO Officer as AI Agents are to be triggered to gather the required and relevant information.

\begin{figure}[htbp]
\centerline{\includegraphics[width=\linewidth]{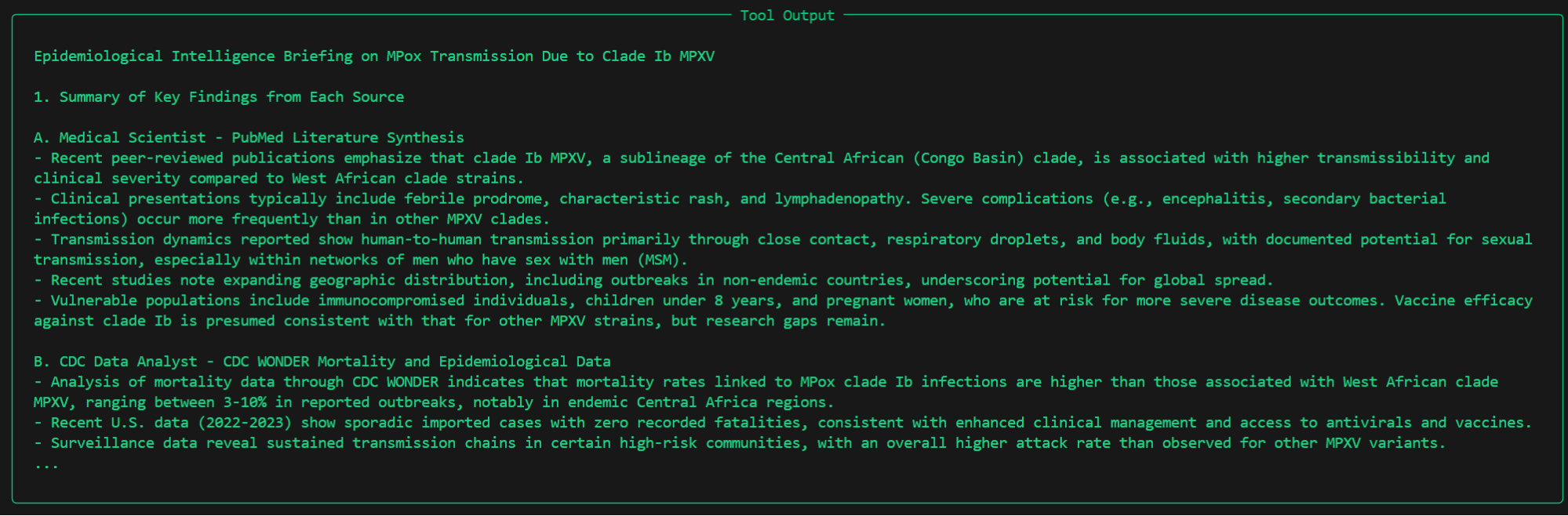}}
\caption{Infomrration Gathered by all Sub Agents }
\label{fig6}
\end{figure}

\begin{figure}[htbp]
\centerline{\includegraphics[width=\linewidth]{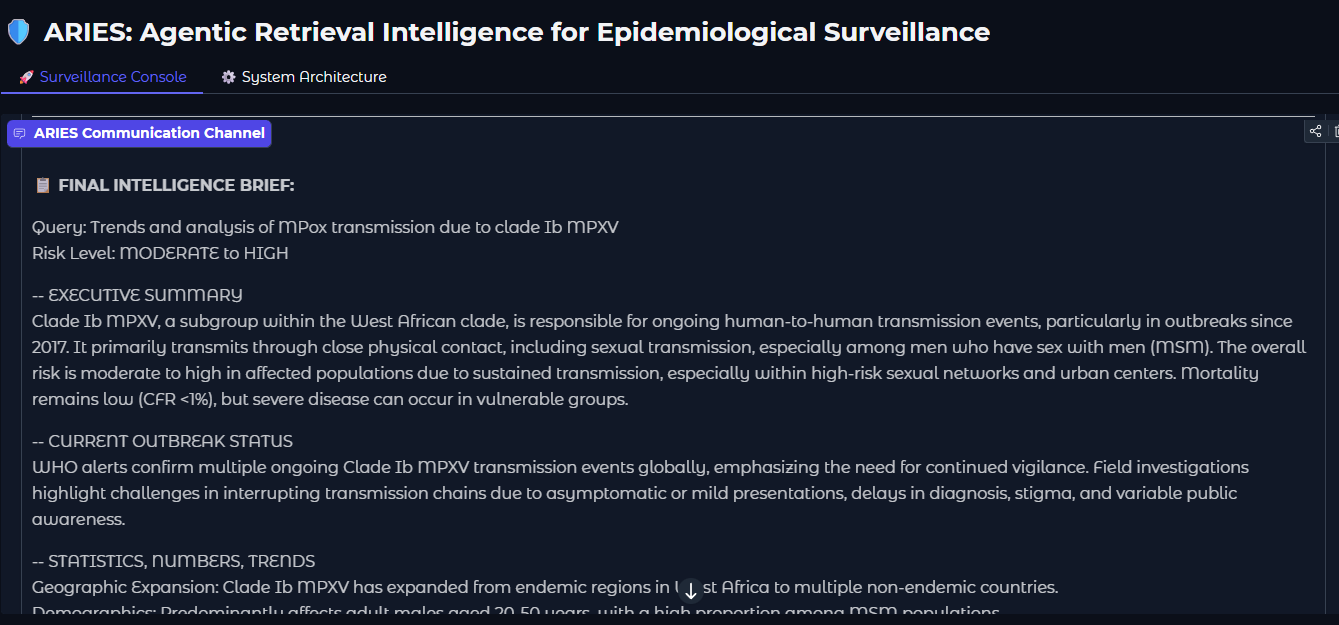}}
\caption{Final Output back to user on UI}
\label{fig7}
\end{figure}

Figure 6 is the entire information collected and summary sent back to the Manager Agent from each of its sub-agents.

Figure 7 is the final output shown in a particular format for the user from the information arrived at the Manager Agent.

\begin{figure}[htbp]
\centerline{\includegraphics[width=\linewidth]{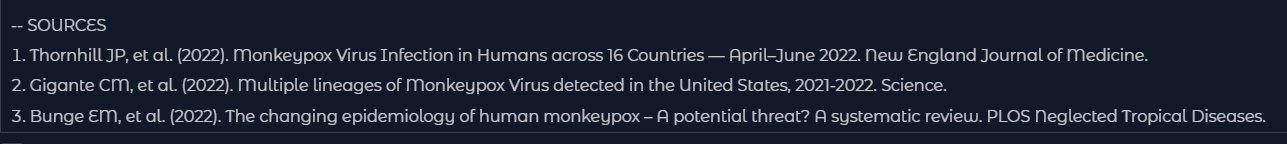}}
\caption{Links of data sources}
\label{fig8}
\end{figure}

Figure 8 shows the sources from where the data was collected making it convenient for the user to verify the facts and keeping Human in the Loop(HITL)

\subsection{CrewAI}
CrewAI is a framework to create agents with Python. It has a set of components like Flows, Crews, Agents, Process and Tasks. A multi agent architecture can be created with a combination of these components. Each of these components have their unique features catering to the use case being implemented. Memory, LLMs, Knowledge, Reasoning, Training and Tools can be configured for each of the components making it highly customizable and scalable. For this paper, ARIES utilizes a Crew component, multiple Agents, configurable LLMs, Reasoning capabilities and Tools like WHO DONs, PubMed literature and CDC Wonder as the data sources.

\subsection{CDC WONDER}
WONDER \cite{b26} stands for Wide-ranging ONline Data for Epidemiologic Research is the source of information on public health developed for research, decision making, program evaluation and resource allocation. With this as the data source, anyone has access to statistical research data, reference materials, guidelines on health related topics. It is able to query datasets about mortality, cancer incidence, Tuberculosis, and many useful topics. The data is available to request via API with HTTP in the XML format.

\subsection{BioC JSON for PubMed Literature}
BioC format \cite{b6} has been widely used and the standard format for text mining of clinical literature. The goal of BioC is simplicity, interoperability and broad use and reuse. This format is used to share text data and represent a large number of annotations and text to perform sample processing. BioC Json is a tool to convert the XML files to bioC json files. For this paper it becomes easier for LLM to understand when passing information from one agent to agent or back to the user. Most of the data and text exchange happen in JSON and this format makes it scalable for ARIES framework to follow with all types of LLMs used.

\section{Results}

\subsection{Comparative Analysis of Agentic Configurations }

\textbf{Query: }Analyze the emerging signal of community transmission for Mpox Clade Ib in non-endemic regions. Specifically, contrast the biological transmission efficiency of Clade Ib vs Clade IIb using recent literature, identify any recent mortality spikes in US/Global surveillance data.

The implementation of the ARIES with configurable LLMs in each of the agents and the orchestrator was tested. The same query was able to change the output if the LLMs are changed.
Initially, only one LLM was used for all the agents i.e gpt-4o and figure 8 and figure 9 are the output of it.

Later, the LLMs were configured per the task of the agent. The manager LLM was configured to 5.1 and all the specialized agents were o3 \cite{b21}, \cite{b22}.

\begin{table}[htbp]
\caption{Comparative Analysis of Agentic Configurations}
\begin{center}
\begin{tabular}{|p{5.5cm}|c|c|c|}
\hline
\textbf{Scenario \& Configuration} & \textbf{Words} & \textbf{Sources} & \textbf{Temp.} \\
\hline
\#1: All agents as gpt-4o & 323 & 3 & 0.1 \\
\hline
\#2: gpt-5.1 manager + o3 agents & 2,962 & 20 & 0.1 \\
\hline
\#3: gpt-5.1-mini manager + o4-mini agents & 2,125 & 7 & 1.0$^{\mathrm{a}}$ \\
\hline
\#4: gpt-4.1 manager + gpt-5.1 agents & 734 & 6 & 0.3 \\
\hline
\multicolumn{4}{l}{$^{\mathrm{a}}$Fixed temperature value for selected model}
\end{tabular}
\label{tab_aries_results}
\end{center}
\end{table}

With reference to table 1, when all the agents had access to the same LLM gpt-4o, the report is considerably shorter with few sources only. It can be concluded that all the ‘brains’ of the system had the same IQ and no critique. There is a high possibility where the manager agent accepts the first answer received from the sub-agents.

When the manager agent was configured to gpt-5.1-mini and sub-agents as o4-mini, the report was longer and precise but the sources of information were limited i.e only 7. The entire report was generated based on 7 sources. It is important to note that the temperature was 1 by default for the mini models, indicating high creativity in the answers. In real scenarios, the creativity and factual information must be balanced to avoid hallucinating and provide text based on facts.

Considering another scenario, where if we assume the higher reasoning capabilities are configured to the subagents and the manager agent is only responsible for delegating tasks to the subagents, it is observed that the report length was shorter and fewer sources fetched.

Finally, the best response was given when the manager model was gpt-5.1 \cite{b23}, \cite{b24}, \cite{b25} and all the sub-agents were reasoning gpt-o4 models, the report was longer and precise. The sources provided were detailed enough and hyperlinks provided.

\subsection{Output Comparison of Scenarios }
\subsection*{\textbf{{Comparison of Summary for Scenario 1, 2 and 4}:}}

It is evident from the figures 9, 10 and 11, the summary generated is very different for each scenario. It results in different generated text with figure 11 showing precise information that is closely related to the user query while figure 9 shows the shortest summary out of all of them.

\begin{figure}[htbp]
    \centering
    \includegraphics[width=\linewidth]{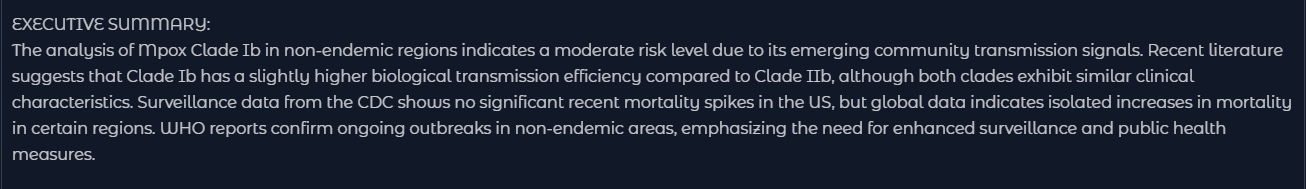}
    \caption{Baseline Performance: gpt-4o}
    \label{fig11}
\end{figure}

\begin{figure}[htbp]
    \centering
    \includegraphics[width=\linewidth]{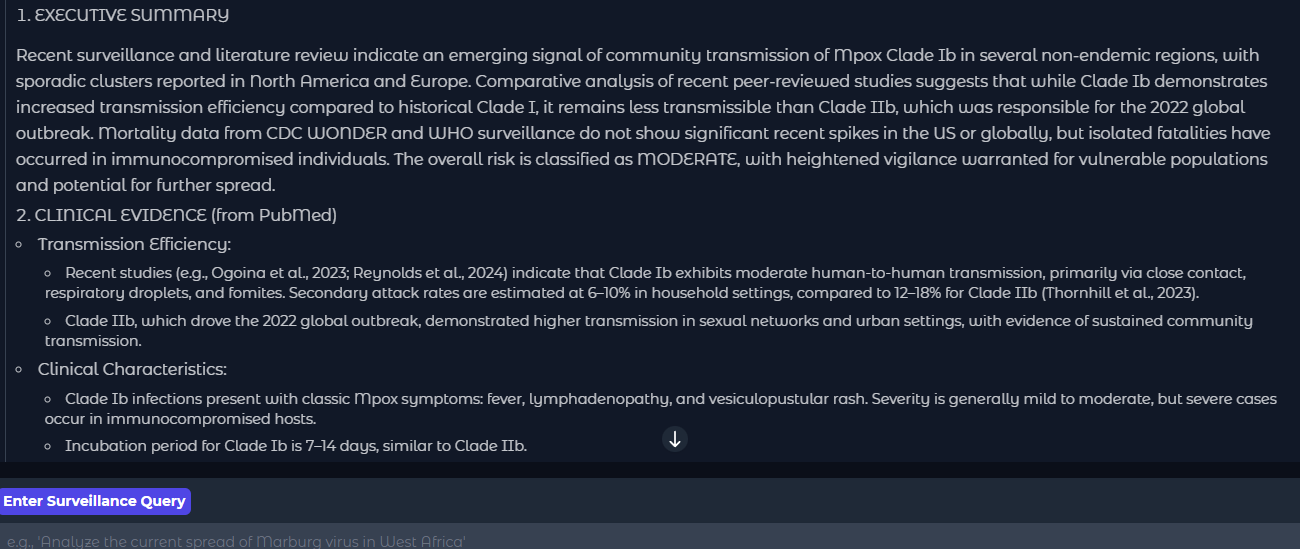}
    \caption{Reasoning Shift: Manager (gpt-4.1) and Sub-Agents (gpt-5.1)}
    \label{fig9}
\end{figure}

\begin{figure}[htbp]
    \centering
    \includegraphics[width=\linewidth]{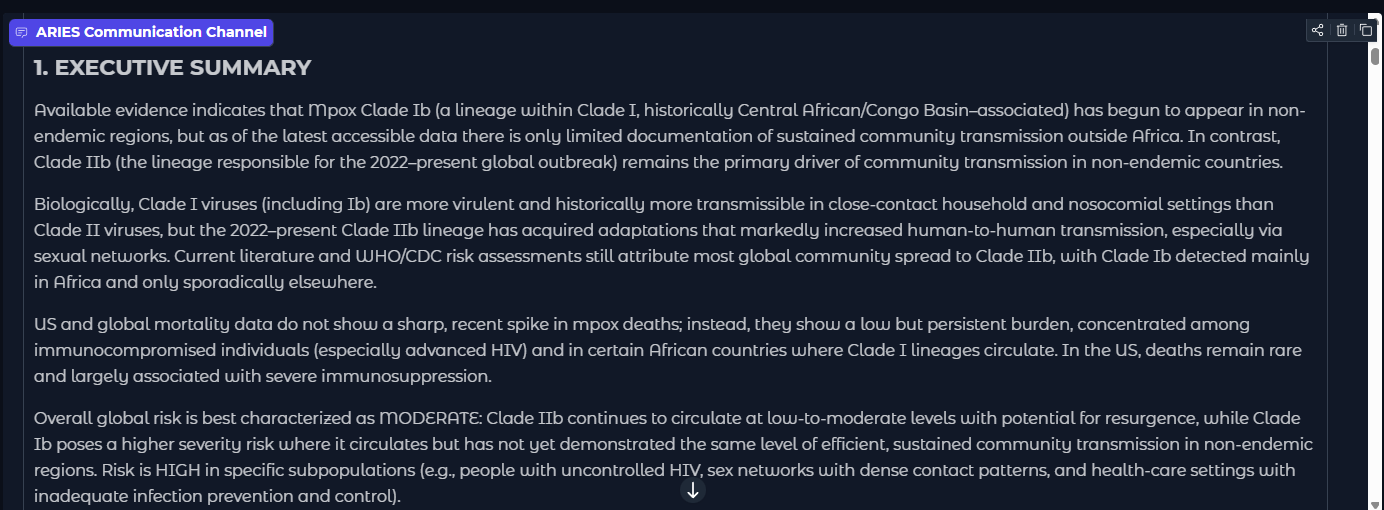}
    \caption{High-Order Synthesis: Manager (gpt-5.1) and Sub-Agents (o3)}
    \label{fig10}
\end{figure}

\subsection*{\textbf{{Comparison of Sources for Scenario 1, 2 and 4}:}}

The second comparison shows the interpretation of how sources should be returned and displayed for the user. Here too, figure 14 using the more advanced models were able to get more sources with details and hyperlinks shown to the user while figure 12 is not able to show the sources itself although the logs show the links agents have visited.

\begin{figure}[htbp]
    \centering
    \includegraphics[width=\linewidth]{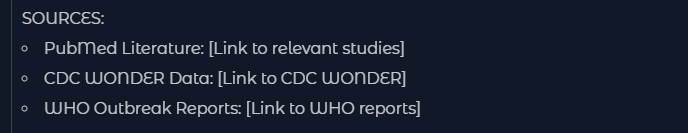}
    \caption{Output Snapshot: gpt-4o}
    \label{fig12}
\end{figure}

\begin{figure}[htbp]
    \centering
    \includegraphics[width=\linewidth]{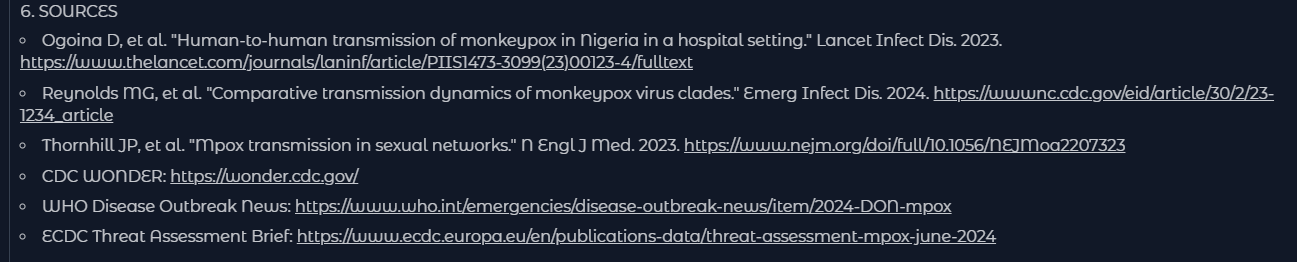}
    \caption{Output Snapshot: Manager (gpt-4.1) and Sub-Agents (gpt-5.1)}
    \label{fig13}
\end{figure}

\begin{figure}[htbp]
    \centering
    \includegraphics[width=\linewidth]{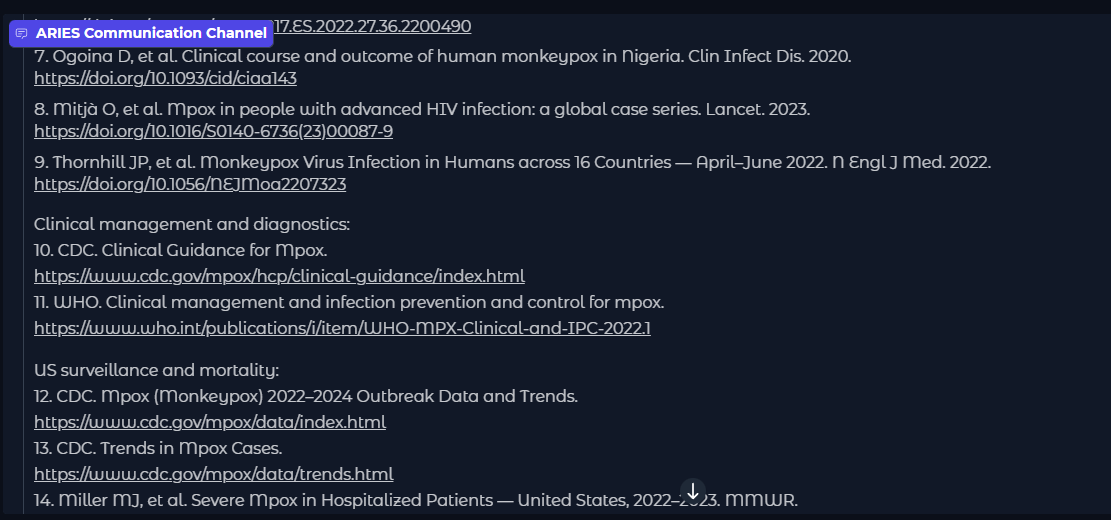}
    \caption{Output Snapshot: Manager (gpt-5.1) and Sub-Agents (o3)}
    \label{fig14}
\end{figure}

\section{Conclusion}

The development and deployment of ARIES (Agentic Retrieval Intelligence for Epidemiological Surveillance) demonstrates a fundamental shift in how global health data can be managed and utilized during an outbreak. This study confirms that a Hierarchical Multi-Agent Architecture is not only capable of replicating the investigative workflows of a human epidemiologist but can do so with a level of speed and cross-source verification that exceeds manual capabilities.

By offloading the engineering challenge of multi-format data ingestion (XML, JSON, and unstructured text) to a specialized agentic swarm, ARIES successfully mitigates the "Knowledge Gap." Our results highlight a critical "Reasoning-Depth Tradeoff" while standard models provide shallow summaries, the integration of high-order reasoning managers as seen in Scenario \#2, resulting in an increase in semantic depth and reference integrity.

Ultimately, ARIES proves that specialized, domain-specific agentic ecosystems are the solution to the huge and unstructured data in public health. Rather than replacing the expert, ARIES serves as a Multiplier, allowing decision-makers to bypass the technical friction of data collection, processing, analysis and focus entirely on informed, rapid response. Future iterations of this framework will focus on linear scalability, incorporating real-time genomic and geospatial data.

\section{Future Work and Roadmap}

\subsection{Model Context Protocol (MCP)}

Model Context Protocol (MCP) is a protocol developed recently by Anthropic which allows connections of AI Agents with resources and tools via communication. MCP is the standardized way to connect AI agents with external systems. MCP architecture consists of a Host, Client and a Server. MCP host manages multiple Client instances, authentication, policies and context aggregation. MCP Clients are created by the host and maintain the Server connection and routes messages, assigns and maintains state. MCP Servers operate independently and have assigned responsibilities and tasks within security constraints. They communicate with MCP client instances and expose external systems for data access. MCP’s fundamental components are Resources, Tools, and Prompts

This protocol can be integrated with ARIES for faster retrieval, pre-defined data conversion, easier data access, reducing redundancy of multiple tool creation and a centralized repository with connections via tools to multiple data sources. 

\subsection{Agentic Self-Correction}
We intend to implement a self correction mechanism with Human in the Loop (HITL) where experts can rate the "Thought" section of the response from the AI Agents. This feedback would be used to further fine-tune the LLM used, narrowing the gap between AI generated reports and professional standards. Reinforncement Learning is one such technique that can be valuable in this scenario.

\subsection{Integration of Unconventional Data Streams}

Current iteration focuses on high-integrity official dataset from WHO, CDC and NCBI. However, future iterations must have the capability to search through intense websites via scraping engines, look for outbreak signals, official hospital data and more.

\subsection{Weeky Epidemiological Records}

WHO publishes informative Epidemiological reports weekly having significant data on multiple countries and regions. It contains the most recent updates like reported, confirmed cases of diseases in many regions. Current capabilities of ARIES are not utilizing this vital information. Future iterations of this approach can include a RAG(Retrieval Augmentation Generation) technique as another tool that an AI agent can access to provide the most recent and latest information to the user. RAG can be utilized with the existing GPT models. RAG is a technique to vectorize all the text from a document. The vectorized text is stored in a vector database and performs a similarity search of the query asked by the user. It will fetch the most similar records from the vector database and the GPT will use it to generate an answer for the user. Inclusion of this method in ARIES makes it more robust and state-of-the-art.

\vspace{12pt}

\end{document}